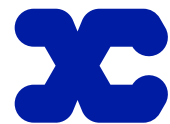

# Gearbox designs for the diamond anvil cell: Applications to hard-to-reach-places

Paul B. Ellison,[1]* Dean Smith,[2] Stanislav Sinogeikin,[3] & Ashkan Salamat[2,1]

1.  Department of Physics & Astronomy, University of Nevada, Las Vegas, Las Vegas, NV  89154, USA
2.  Nevada Extreme Conditions Laboratory, University of Nevada, Las Vegas, Las Vegas, NV  89154, USA
3.  DAC Tools, LLC., Naperville, IL  60565, USA

The uniaxial compression design of the diamond anvil high pressure cell (DAC) necessitates the use of a soft pressure-transmitting medium (PTM) to minimize non-hydrostatic effects at significantly high pressures, and many such media are gaseous at ambient conditions. There now exist a number of commercially-available instruments – high-pressure gas loading apparatus – for the insertion of gases into the diamond anvil cell up to several kbar. These machines allow the use of gases as a PTM, as a reagent for high-pressure chemistry, or to be loaded as the sample material itself. We present the development of two highly adaptable gearbox designs to allow the controlled closure of the DAC within an atmosphere of gas at several kbar, contained within the walls of a pressure vessel. The first applies a torque directly to the pressure screws and is thus DAC-specific, while the second applies a load to the body of the cell and may be used with a wide variety of DAC designs.

A myriad diamond anvil cell (DAC) designs exist, often tailored to specific measurements. In many experiments, there is a necessity to load materials into the DAC which are gaseous at ambient conditions. These experiments may be in study of the gas itself under extreme conditions,[1,2] may utilize the gas as a reagent for a novel chemical reaction only accessible at extreme pressures,[3,4] or benefit from the use of the gas as a pressure-transmitting medium (PTM) – a material which remains soft into the solid phase and encapsulates the sample, acting to reduce non-hydrostatic stresses in the sample and compensating for the uniaxial compression inherent to the design of the DAC.[5–8] A common method towards the insertion of these gases into the DAC is the use of a high-pressure gas loading apparatus, in which the DAC is contained within a vessel and pressurized to the kbar regime whilst open, and the subsequent closure of the DAC in this environment traps high-pressure gas within the DAC sample chamber. This provides a high starting gas density for the experiment, before any load is applied by the diamonds. Crucially, this avoids a significant portion of the volume collapse in the gas by having the experiment begin some distance along its P–V equation of state, thereby reducing the amount of deformation on the sample chamber throughout the high-pressure experiment.

The closure of the DAC within the high-pressure gas loading apparatus thus presents an engineering challenge – how does one apply a controlled load to the DAC when confined within the thick (typically several inches) walls of a pressure vessel? Here, we describe the design of two gearboxes tailored to the closure of a diamond anvil high-pressure cell during the gas loading procedure: One design that mates with a specific DAC geometry, and a second which may be used to close DACs of various designs.

## DAC-specific gearbox

The gearbox design is shown in **Figure 1a** and photographs in **Figure A1**, with its geometry designed for that of our in-house "PEAS-Q36" DAC design. A pair of hex sockets on the top of the device drive two pairs of hex drivers on the bottom, which apply torque to the DAC pressure screws. The hex sockets on the top of the device are aligned to match the mechanical feedthroughs from the bulkhead of a high-pressure gas loading apparatus.[9] The gears consist of 2 drive gears, which are manually turned through the bulkhead, and 4 driven gears which turn the DAC pressure screws. The gearing is configured at 1:1 ratio. The 4 driven gears are sprung, allowing the device to switch between two modes – free-spinning and engaged. When not in contact with the DAC pressure screws, springs push the 4 driven gears out of engagement from the 2 drive gears. This also causes the hex drivers to greatly protrude from the face of the gear box, providing visual confirmation of disengagement.

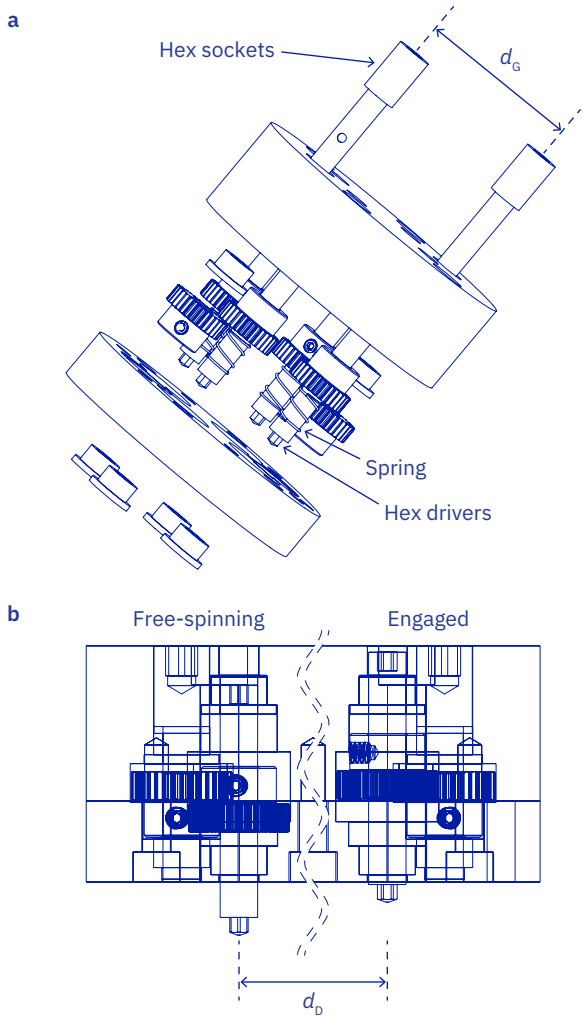

**Figure 1.** Construction and operation of gearbox for closing diamond anvil cell. (a) Exploded view of CAD model showing the interior gearing of the device. (b) Modes of operation of the gearbox, in free-spinning mode the user can align the extruding hex drivers to the DAC pressure screws, and then enter engaged mode.

**Figure 1b** shows the two modes of operation of the gearbox design. In free-spinning mode, the extruding hex drivers (attached to the driven gears) may be manually turned to align each with the pressure screws on the DAC. Once each hex driver is aligned with the DAC pressure screws, the gearbox may be pushed down to engage the gears and seal the containment vessel. The addition of free-spinning mode thus allows ease of alignment of the gearbox for quick assembly of the gearbox and container, allowing the positions of the DAC pressure screws to remain untouched throughout the assembly procedure – an often crucial consideration, for instance in the case of an experiment concerning an air-sensitive sample in a hermetically-sealed DAC. Furthermore, the travel on the gearbox between free-spinning and engaged modes ensures that the containment may not be closed without each hex driver engaged with a pressure screw.

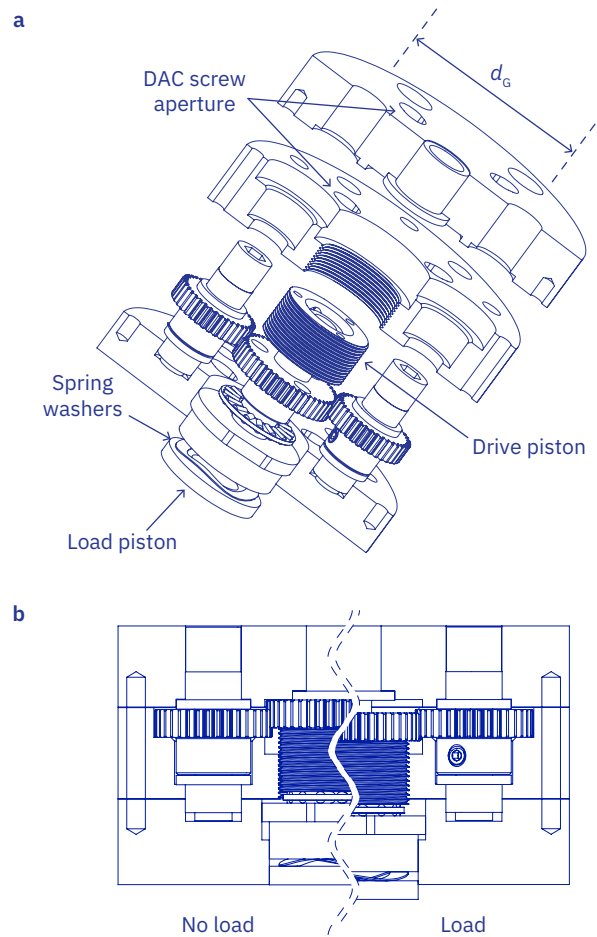

**Figure 2.** Construction and operation of universal gearbox for closing a diamond anvil cell. (a) Exploded view of CAD model, (b) Operation of universal gearbox, showing action of top screws to drive the piston.

The dimensions of the gearbox are defined by both the pressure vessel of the gas loading apparatus and the design of DAC via two parameters: (i) distance $d_G$ separating the driver feedthroughs in the bulkhead which mate with the gearbox (in our case, $d_G = 1.75$"), and (ii) the position of the DAC pressure screws, separated by a distance $d_D$. The pressure screw hex driver requirement defines only the size of extruding hex driver, however, the handedness of the screws has significant implications for modifications to the drive configuration – note that many DAC designs implement left-handed pressure screws to facilitate comfortable turning of two screws at once by hand. At present, gearboxes of our design have been constructed for compatibility with the PEAS and the BX-90 DAC, both of which feature only right-handed screws. Provided that accommodations are made for these constraints, the present design can be modified for compatibility with a wide variety of screw-driven DAC designs.

## Universal gearbox

Our design for a universal gearbox is shown in **Figure 2**, and photographs of the device are shown in **Figure A2**. The universal gearbox utilizes the same drive gear configuration as the DAC-specific gearbox, and $d_G$ is the same.

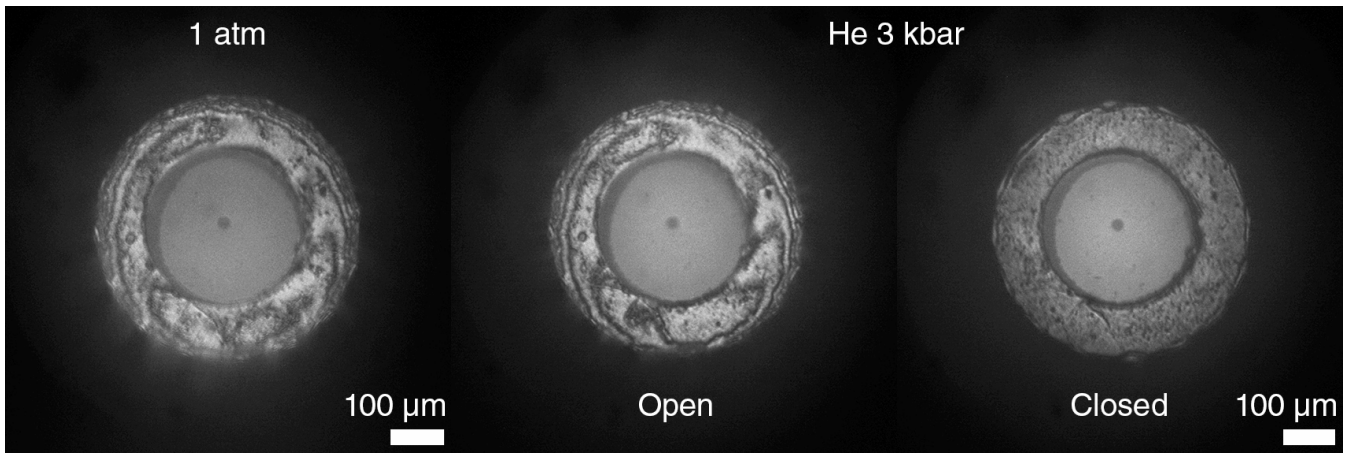

**Figure 3.** Closure of a Princeton-type symmetric DAC by the universal gearbox during high-pressure gas loading of He at 3 kbar. The DAC is fitted with diamonds with 500 μm diameter culets, and a stainless steel gasket with a 280 μm diameter sample chamber containing a ruby sphere. The disappearance of interference fringes between the gasket material and diamond signify the closure of the DAC.

In this design, rather than turning the DAC pressure screws by geared transfer, the gears turn a finely-threaded center piston – the drive piston. The drive piston then transfers axial force to a secondary, non-rotating piston – the load piston. A thrust needle roller bearing is placed between the rotating drive piston and load piston to minimize friction and rotational forces. Precisely fitted flats are machined into the housing and onto the load piston in order to prevent its rotation.

The load piston contains a set of spring washers, dampening the load on the DAC for a given torque on the drive piston, thus providing the user with more control over the applied load. The spring washers also assist in maintaining closure of the DAC during release of gas from the gas loading apparatus, serving as a buffer to the differences in material compression between the aluminum and steel components.

An important difference to note between the DAC-specific gearbox and the universal design is the mechanism for applied load. In the DAC-specific gearbox, the DAC is closed by turning driven gears which in turn, tighten the DAC pressure screws. Thus, the DAC will remain closed on extraction from the gearbox and its containment vessel. In the universal design, load is applied by direct force onto the body of the DAC applied by the gearbox piston and the containment, and the closing load must be transferred to the DAC pressure screws following extraction from the gas loading apparatus, and prior to removal from the containment. To facilitate this, apertures through the universal gearbox allow access to the DAC pressure screws. As the load from the gearbox is applied by a single piston along the central axis, *i.e.* it is not sensitive to the orientation of the DAC about this axis, the DAC may be oriented inside the containment such that its pressure screws are aligned with the apertures in the universal gearbox. Additional apertures along the central axis permit optical access to the DAC through the body of the gearbox as well as the base of the containment vessel, allowing inspection by microscope or optical measurements (*e.g.* pressure measurements by ruby photoluminescence or Raman spectroscopy). Users may then transfer the load onto the DAC pressure screws controllably before opening the containment vessel to extract the prepared DAC.

For *in situ* imaging during the gas loading procedure, we recently described an optical microscope which can form high-quality images even following the significant (up to 20% [10,11]) change in refractive index of the gas arising from compression.[9] There, the imaging lens is placed directly onto the PEAS-Q36 DAC. As the universal gearbox may be used with DACS of a multitude of designs, we instead opt to have a threaded plug to house the imaging lens, and the aperture at the base of the containment vessel accepts the threaded plug. In this way, lenses may be easily exchangeable to suit the focal length requirements of a variety of DACs, and the plug may be removed after extraction from the gas loading apparatus to allow the aforementioned optical access.

In principle, the present gearbox design can work with diamond anvil cells of almost any geometry, provided a containment vessel of appropriate dimensions is created.

**Closure of a DAC in high-pressure gas-loading apparatus**

Crucial to repeatable, successful closure of the DAC in the high-pressure gas vessel is reliable imaging of the sample environment. As described above, we have modified our previously-described design to accommodate a variety of DAC geometries, and here apply the principle to the symmetric "Princeton" design.[12] A ZnSe lens with focal length 22.6 mm at $\lambda = 633$ nm is selected as the imaging lens. As before, ZnSe is used for its high refractive index, minimizing the focal length change as the gas density increases during loading.

**Figure 3** shows optical images of a gasket and sample chamber formed from inside the gas loading apparatus throughout the loading procedure. At 1 bar and at 3 kbar, the cell is confirmed to be open by the presence of interference fringes between the diamond culet and metal gasket. Closure of the DAC is performed by applying load with the universal gearbox design onto the body of the DAC, and is applied until interference fringes are no longer visible. In this way, the DAC is closed at the gas pressure of the loading apparatus – in this case 0.3 GPa. The sample chamber in Figure 5 contains a ruby sphere. On extraction from the gas loading apparatus, the load is transferred to the DAC by turning the pressure screws, and removing load from the piston of the universal gearbox.

## Conclusions

We present two designs of gearbox designed for the closure of a diamond anvil high-pressure cell in hard-to-reach places – notably the vessel of a high-pressure gas loading apparatus. The DAC-specific gearbox may be adapted to work with screw-driven DAC designs of any dimensions, provided suitable considerations are made. The universal gearbox may be used with a wide variety of DACs in its current form, provided that an appropriate containment vessel is available.

## Acknowledgements

The authors thank David P. Shelton.

# Appendix: Photographs

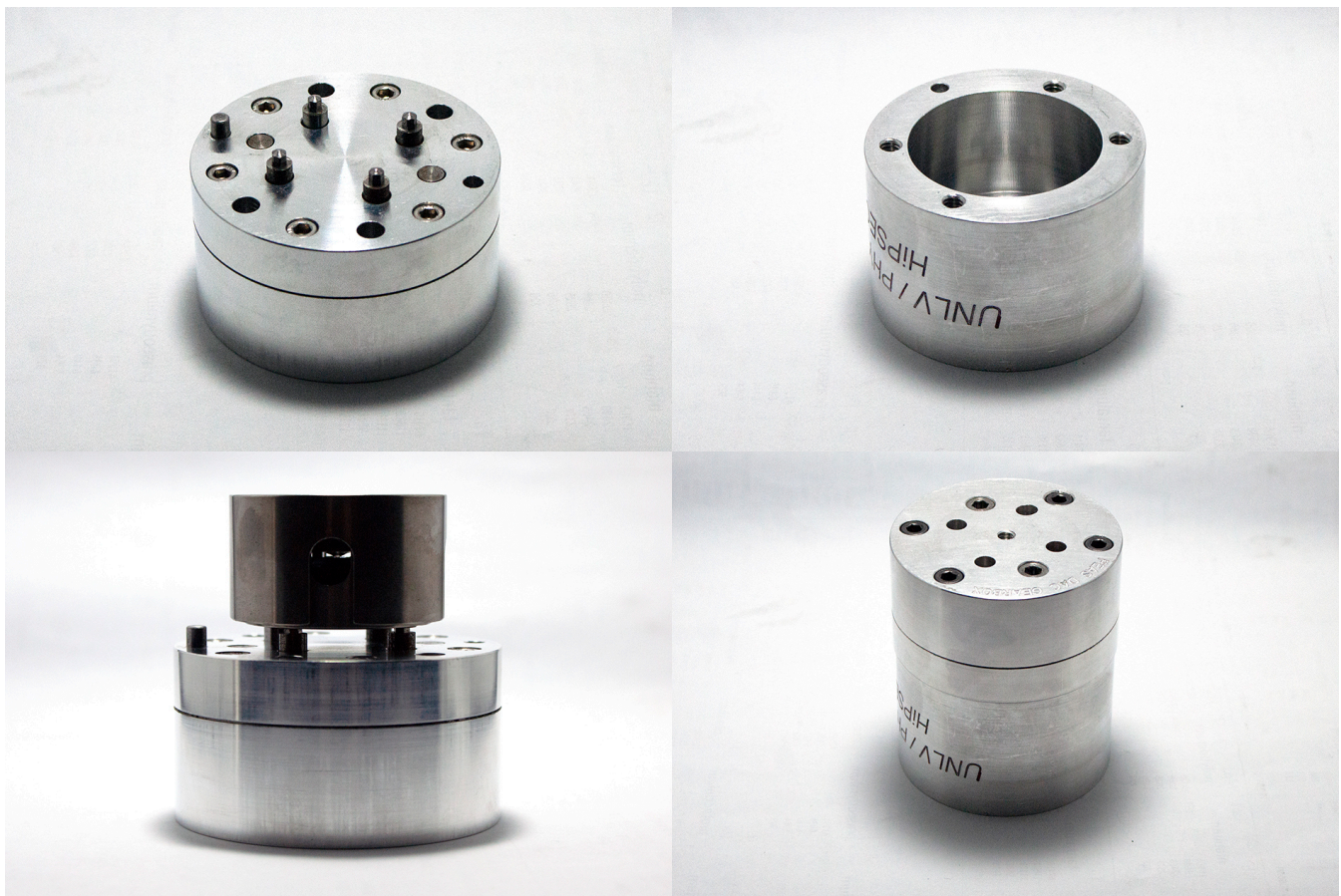

**Figure A1.** Photographs showing the assembly of gearbox for closing PEAS-Q36 DAC design. **(top-left)** Gearbox design with four extruding hex drivers. **(top-right)** Example of containment vessel for DAC. **(bottom-left)** A PEAS-Q36 DAC is mated with the gearbox. Note that the hex drivers still extrude from the gearbox. **(bottom-right)** The mated DAC and gearbox are inserted into the containment vessel, and driven closed by four pressure screws. In closing the assembly, the gears within the gearbox are brought into engaged mode.

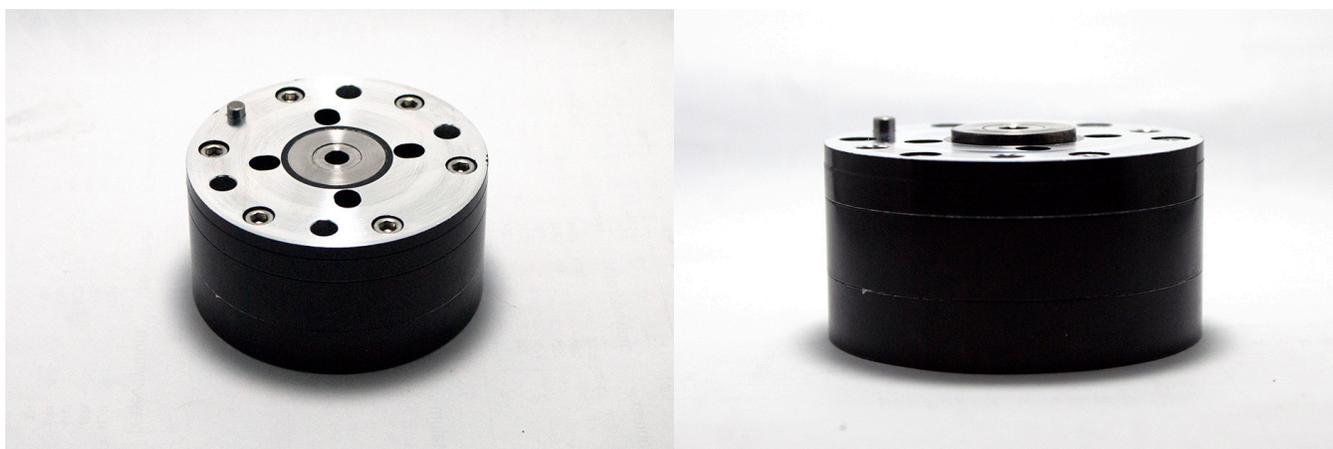

**Figure A2.** The universal gearbox for closing DACs of various design. **(left)** The load piston remains flush with the bottom of the gearbox during assembly. **(right)** On turning the hex sockets on the top face of the gearbox, the load piston extrudes from the gearbox to exert force onto the DAC.